\documentclass{ws-procs975x65}
\usepackage{textcomp}

\begin{document}

\title{SEARCH FOR NEUTRINO OSCILLATIONS \\ IN APPEARANCE MODE \\ WITH THE OPERA EXPERIMENT}

\author{T. A. DZHATDOEV$^*$ on behalf of the OPERA Collaboration and V. L. TROSHINA}

\address{Skobeltsyn Institute of Nuclear Physics, Lomonosov Moscow State University,\\
Moscow, Russia\\
$^*$E-mail: timur1606@gmail.com}

\begin{abstract}
The present paper highlights the data analysis status of the OPERA experiment. The experiment was designed to perform the neutrino interactions analysis on event-by-event basis, and optimized to search for $\nu_{\mu} \rightarrow \nu_{\tau}$ oscillation in appearance mode, also allowing to perform a $\nu_{e}$ appearance search.

We review the data simulation and the analysis chains implemented to search for $\nu_{\tau}$ interactions. The main kinematical parameters sensitive to the neutrino flavor are discussed, the uncertainties of the event parameter estimation are reviewed, and the main sources of background for the $\nu_{\mu} \rightarrow \nu_{\tau}$ oscillation search are examined. The topologies of the two first $\nu_{\tau}$ candidate events are presented.

Finally, we review the status of the $\nu_{e}$ appearance search and present the constraints set by the OPERA experiment on the mixing angle $\theta_{13}$ and on the LSND/MiniBooNE anomaly.
\end{abstract}


\bodymatter

\section{Introduction} \label{sec:intr}
An idea of neutrino oscillations was introduced by B.M. Pontecorvo long ago \cite{pon57}. However, it took over 40 years to get the first definite experimental result in favor of oscillations \cite{fuk98}. Now it is firmly established that the number of the ``active'' neutrino flavors amounts to 3 (formal fit gives 2.998$\pm$0.008 \cite{hag02}), thus oscillations between these states are described by 3$\times$3 Pontecorvo-Maki-Nakagawa-Sakata (PMNS) matrix \cite{mak62}.

Further experiments with solar \cite{ahm02}, atmospheric \cite{amb98,all05}, reactor \cite{ara05} and accelerator \cite{ahn06,mih06,ada08} neutrinos gave additional support to the oscillation hypothesis. Recently the value of the last previously unknown mixing angle $\theta_{13}$ was measured in the Daya Bay ex\-pe\-ri\-ment \cite{an12}. Some phenomenological models introduce and some experiments (e.g. \cite{aqu01,aqu12}) support the existence of $m$ additional (``sterile'') neutrino states that may participate in oscillations; in that case the PMNS matrix is $3+m$-dimensional.

Other unsolved problems, besides the question of sterile neutrinos existence, are those of neutrino mass hierarchy and of CP violation in the leptonic sector. The decisive proof of $\nu_{\tau}$ appearance in $\nu_{\mu} \rightarrow \nu_{\tau}$ oscillation is also still absent; only recently the Super-Kamiokande experiment found the evidence for this effect \cite{abe13}.

The OPERA experiment was designed to search for $\nu_{\mu} \rightarrow \nu_{\tau}$ oscillation on event-by-event basis by observing $\tau$ decays. OPERA might be also handy in constraining the active-sterile neutrino mixing parameters, and the so-called non-standard interaction (NSI) parameters \cite{ter11}, that may give rise to $\nu_{\tau}$ component in the initially pure $\nu_{\mu}$ beam.

In the present paper we review the data analysis status and the results of the OPERA experiment. Much more detailed description of the data analysis chain is given in \cite{aga13a}. For the OPERA hardware reference see \cite{acq09}. The status of emulsion scanning is presented in the other OPERA contribution in this volume \cite{she13}.

In sec.~\ref{sec:chain} we present a sketch of the OPERA analysis procedure. In the following sections some details of simulations and analysis are given: the parameters sensitive to the primary neutrino flavor are introduced in sec.~\ref{sec:param}, the emulsion detector capabilities to reconstruct the neutrino interaction kinematical parameters are discussed in sec.~\ref{sec:est}, the background studies are summarized in sec.~\ref{sec:bckg}. In sec.~\ref{sec:cand} we present the two first $\nu_{\tau}$ candidates, and in sec.~\ref{sec:elec} --- the results concerning the $\nu_{e}$ appearance in the OPERA experiment.

\section{The sketch of the OPERA simulation/analysis chain} \label{sec:chain}
OPERA is a very complex hybrid detector that consists of electronic and emulsion detectors that are able to sample the event development on highly different spatial scales, starting from $L \approx 20$ $m$ --- the lenght of the full detector --- to sub-micron level needed to sucessfully identify the $\nu_{\tau}$ events. OPERA DAQ \cite{mar10} has a high efficiency of response to neutrino interactions contained in the detector volume (according to \cite{aga11}, more than 95 \%). However, most of $\nu_{\tau}$ signal gets lost at the further stages of the event reconstruction and analysis, and the final $\nu_{\tau}$ finding efficiency is about 6-8 \%.

The main steps of the standard OPERA event analysis chain are:
\begin{arabiclist}[(6)]
\item Selection of contained events that have the interaction (primary) vertex inside the OPERA volume \cite{ber09}.
\item Search for the element of the emulsion detector --- the so-called brick --- that contains the primary vertex (brick finding (BF) procedure) \cite{chu13}
\item Each brick contains additional attached emulsion layers (the changeable sheets or CS) needed to confirm the presence of charged particle tracks crossing the brick. The CS are searched for converging patterns or tracks that have an angle close to the reconstructed one from the electronic detector (ED) information \cite{ano08} (see also \cite{aga13a}). In case of the positive signal the corresponding brick is analyzed.
\item Vertex location inside the brick and the event topology reconstruction.
\item Then follows the decay search (DS) procedure needed to search for the signature of the short-lived $\tau$ decay to identify $\nu_{\tau}$ charged current (CC) interactions \cite{ari11}.
\item After reconstruction of the event kinematical parameters, the kinematical selection is applied in order to further supress the background.
\end{arabiclist}

Relevant efficiencies on the different stages of analysis are defined in \cite{gul00} and revised in \cite{aga12a}; their up-to date values are discussed in \cite{aga13a}. Most efficiencies were estimated with dedicated MC simulations. The detector response simulation was performed using the recently introduced software framework OpRelease.

\section{Parameters sensitive to the neutrino flavor in the OPERA experiment} \label{sec:param}

Due to the excellent spatial resolution of the OPERA detector, usually a wealth of information about the event is available. Here we list the main parameters that are sensitive to the primary neutrino flavor, thus making it possible to perform a $\nu_{\tau}$ search. Let us remark that the dominating background for $\nu_{\tau}$ observation is charm production in $\nu_{\mu}$-CC interactions (see sec.~\ref{sec:bckg} for more details).
\begin{arabiclist}[(5)]
\item The presence of muon or electron in the primary vertex allows to exclude the $\nu_{\tau}$ hypothesis in favor of the $\nu_{\mu}$-CC or $\nu_{e}$-CC interaction hypothesis, respectively. Muons are usually clearly seen by the ED as long penetrating tracks (efficiency of muon ID with OPERA ED is $\approx$95 \%) \cite{aga11}. In especially interesting events it is possible to follow down the tracks in other emulsion bricks to refine the muon search.
\item Even in the case when the muon or electron in the primary vertex was missed, efficient selection of $\nu_{\tau}$ is still possible. It appears that the azimuthal angle $\phi$ between the $\tau$ candidate and hadronic jet is a powerful background-supressing variable, as the charmed hadron tends to follow the direction of the primary hadronic jet closely (it is just a part of the jet). On the contrary, transverse momentum conservation enforces the most of leptons ($\mu$ in $\nu_{\mu}$-CC events and $\tau$ in $\nu_{\tau}$-CC events) to have $\phi$ values $>\pi/2$. In practice, the two slightly amended versions of the $\phi$ variable definition are usable.
\begin{alphlist}[(b)]
\item The so-called ``exclusive'' $\phi_{E}$. In events with $>1$ particles in the primary jet the track that has the largest $\phi$ is removed and the $\phi$ parameter calculation is repeated to get the $\phi_{E}$ value. This exclusion procedure is intended to remove not-identified $\mu$ while calculating the $\phi_{E}$ value.
\item Even more sophisticated version of the $\phi$ parameter is the-so-called ``modified'' $\phi_{mod}$. In this case the largest $\phi$ track is excluded unless it is identified as a hadron, which allows to avoid excluding of a genuine hadron from the $\phi_{mod}$ calculation. The $\phi_{mod}$ distributions for the $\tau \rightarrow h$ ($1h$) and $\tau \rightarrow 3h$ ($3h$) decay channels are shown in \cite{aga13a} (fig. 12).
\end{alphlist}
\item For the $\tau \rightarrow \mu$ decay channel: If charge sign of $\mu$ from the secondary vertex is negative, then the charm background is greatly reduced. 
\item For the $\tau \rightarrow 3h$ decay channel: $\tau$ decays into charged particles and neutrinos due to lepton charge conservation, while charm decay may be neutrinoless. This affects the kinematical parameters from the secondary (decay) vertex, especially the transverse momentum, and allows to select a part of $\nu_{\tau}$ events.
\item Other parameters (slope of the $\tau$ candidate track with respect to the primary beam direction, total momentum of the charged ``daughters'', 3D angle between the primary jet and the $\tau$ candidate, etc...) are also sensitive to the neutrino flavor. It is possible to introduce multidimensional method of $\nu_{\tau}$ search. This method is currently under study; it might allow to relax the kinematical cuts and ensure more effective rejection of background, thus enhancing the statistical significance of the $\nu_{\tau}$ observation.
\end{arabiclist}

\section{The event parameters estimation with the OPERA emulsion detector} \label{sec:est}
The most of useful kinematical information in OPERA comes from emulsion brick analysis. The two systems for emulsion data analysis are available: SYSAL (System of Salerno) \cite{bos13} and FEDRA (Framework for Emulsion Data Reconstruction and Analysis) \cite{tio06}.

A well known fact is that the emulsion contains numerous grains of ``fog'' \cite{nak06}, thus one needes to establish additional selection criteria to supress the background while linking the short fragments of the tracks in single emulsion layer --- the so-called microtracks. The main linking criteria demand that the microtracks must have good angular coincidence and the distance between straight lines that contain the microtracks is small. It appears that a part of the tracks from low-energy particles exhibits strong scattering in lead interleaved between the emulsion layers, and thus these tracks are not reconstructed.

\begin{figure}[t]
\begin{center}
\psfig{file=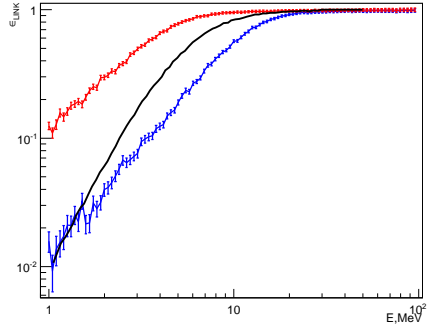,width=2.5in}
\end{center}
\caption{Microtrack linking efficiency vs. energy for electrons (black curve \cite{esp05}); blue and red curves: only angular cut introduced --- 50 and 150 $mrad$, respectively \cite{tro12}.}
\label{fig:fig1}
\end{figure}

The microtrack linking efficiency for electrons vs. energy was calculated in \cite{esp05} and shown in fig.~\ref{fig:fig1} (black curve). Additional studies with the only cut on the angle between the microtracks were performed \cite{tro12} using the Geant4 package \cite{ago03}; their results are shown in fig.~\ref{fig:fig1} by blue (50 $mrad$ cut) and red (150 $mrad$ cut) lines with the statistical uncertainties. Let us remark that the most of hadron tracks are assumed to be reconstructed starting from the minimal momentum of 100-300 $MeV/c$ or even lower.

After the event topology reconstruction one can estimate the particle parameters from emulsion data. Charged hadron momentum measurement with Multiple Coulomb Scattering (MCS) method \cite{aga12b} has an uncertainty $\approx$20 \% for $p$= 4 $GeV/c$ and normal incidence. Gamma or electron energy could be measured after reconstruction of electromagnetic showers with uncertainty $<$30 \% for primary energy $>$ 3 $GeV$ \cite{jug09}. Electron identfication is discussed in \cite{arr07b}, and p/$\pi^{+/-}$ separation in \cite{tos04}.

\section{Background for $\nu_{\tau}$ observation} \label{sec:bckg}
\begin{figure}[t]
\begin{center}
\psfig{file=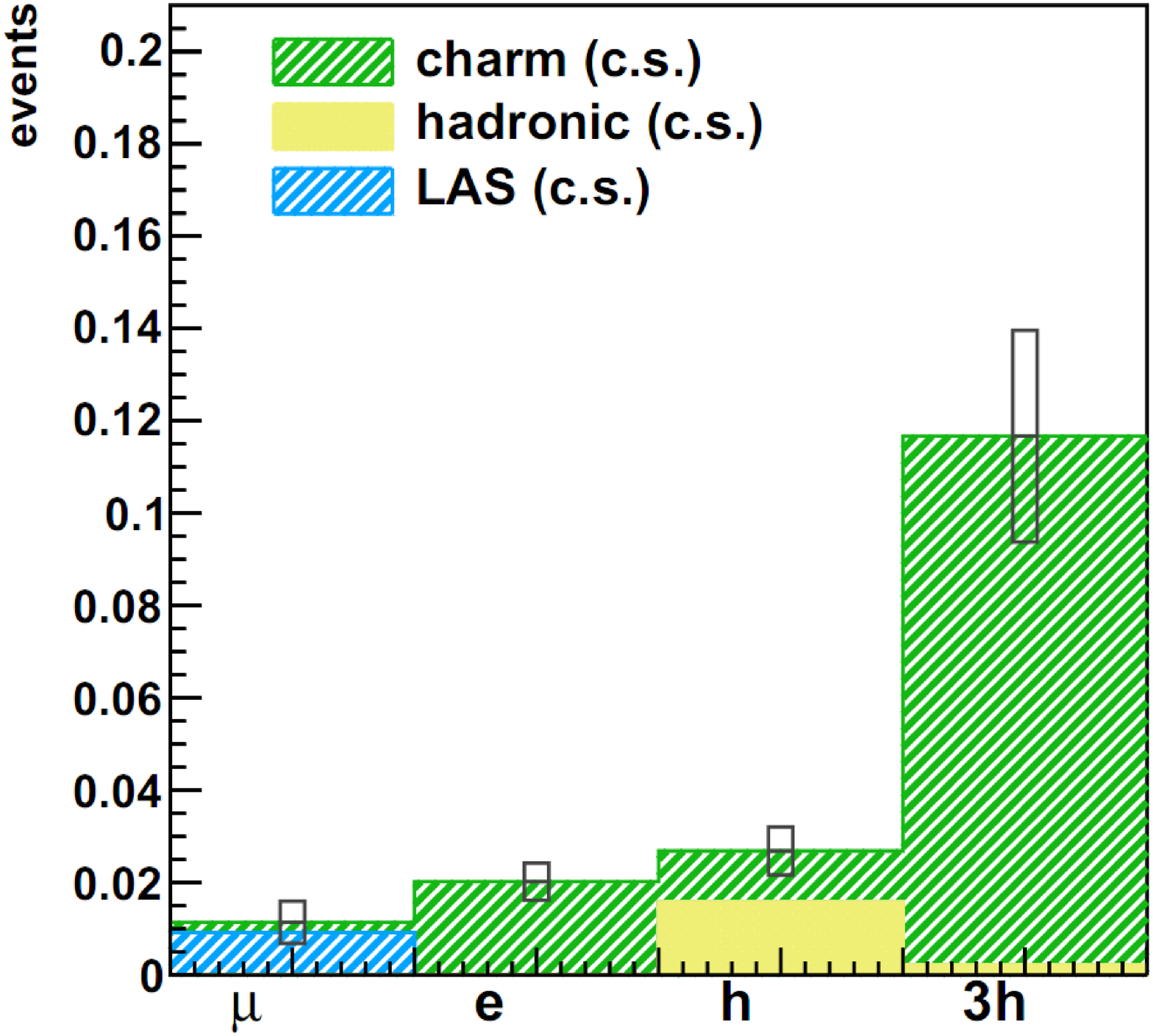,width=2.3in}
\end{center}
\caption{Background expectation for various decay channels \cite{aga13a} (color bars) together with their estimated uncertainties (black rectangles).}
\label{fig:fig2}
\end{figure}

Prompt $\nu_{\tau}$ background is negligible, according to \cite{vdv97}, where actually calculations for the CHORUS experiment were done. The CHORUS beam parameters are comparable with the CNGS parameters. The main sources of background for oscillated $\nu_{\tau}$ observation are as follows:
\begin{arabiclist}[(2)]
\item Background from $\nu_{\mu}$-CC events with charm production. The events with charmed hadron and low-energy muon with large angle ($>$1 rad) w.r.t. the axis of the neutrino beam constitute a very dangerous background. In order to reduce this kind of background, the new methods for large-angle scanning were recently introduced. These methods allow to raise the angular acceptance of scanning microscopes from 0.6--0.8 $rad$ to nearly 1.2--1.3 $rad$.
\item Extensive studies of background from inelastic hadron and muon interactions also were performed (see \cite{aga13a,she13} for more details). 
\end{arabiclist}
Background expectation for various decay channels is shown in fig.~\ref{fig:fig2}. Green bars denote charm background, yellow ones --- background from hadronic interactions; blue bar for the case of the muonic channel stands for large angle muon scattering (LAS). All values are computed for the sample considered in \cite{aga13a} ("c.s." stands for "current sample").

\section{$\nu_{\tau}$ candidate events} \label{sec:cand}
In the sample considered in \cite{aga13a} two $\nu_{\tau}$ candidates were observed, one in the $1h$ and the second in the $3h$ decay channel. In both candidates the short-lived particle track was seen in emulsions (the so-called ``long decays''). The parameters of these candidates are listed in \cite{aga10} and \cite{aga13a} for the first and the second candidate, respectively.

\begin{figure}[t]
\begin{center}
\psfig{file=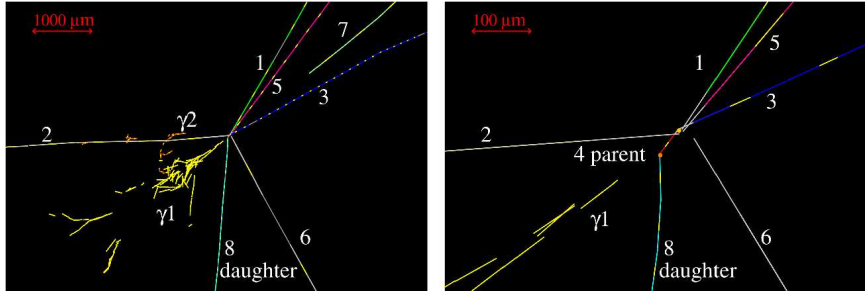,width=4.5in}
\end{center}
\caption{The topology of the first $\nu_{\tau}$ candidate: the area around the vertex is shown.}
\label{fig:fig3}
\end{figure}

The topology of the first candidate in the $1h$ channel is presented in fig.~\ref{fig:fig3}. Here only the region around the vertex is shown; the picture that shows bigger spatial region can be found in \cite{aga10}. Besides the short-lived $\tau$ candidate (``parent''), the event has 5 charged long-lived hadrons from the primary vertex, one track from a neutral prompt hadron that points to the vertex, one charged hadron from decay vertex (``daughter'') and two $\gamma$ showers likely from the decay vertex (only one is shown in fig.~\ref{fig:fig3}). The ``parent'' and the ``daughter'' tracks form a kink. The invariant mass of the system of the 2 $\gamma$ and the ``daughter'' hadron is compatible with the $\rho(770)$ mass.

\begin{figure}[t]
\begin{center}
\psfig{file=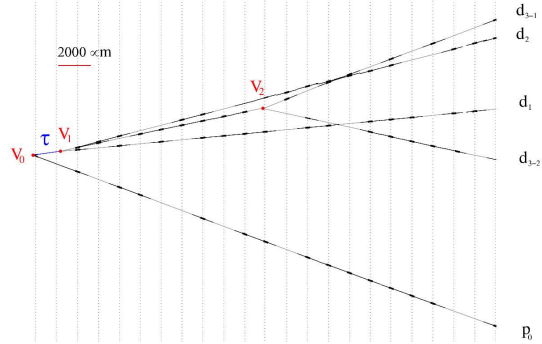,width=2.8in}
\end{center}
\caption{The topology of the second $\nu_{\tau}$ candidate}
\label{fig:fig4}
\end{figure}

Similarly, the topology of the second candidate in the $3h$ channel is shown in fig.~\ref{fig:fig4}. It has the decay vertex in the plastic base. The fact that the charge number of the nuclei that constitute the plastic is much lower that for lead, as well as the density of the material, allows to supress the hadron scattering background. Both candidates don't have muon or electron from the primary vertex, and satisfy all kinematical cuts.

Let us note that recently the third candidate in the $\mu$ decay channel was observed. The publication with more detailed discussion of the third candidate is under preparation \cite{aga13c}.

\section{Constraints on $\theta_{13}$ and search for the LSND/MiniBooNE anomaly with the OPERA experiment} \label{sec:elec}
The CNGS beam contains a non-negligible $\nu_{e}$ prompt component ($\approx$0.8 \% in terms of interactions), thus the OPERA experiment is not optimized to study $\nu_{e}$ appearance. Neutrino energy estimation needed for oscillation analysis was performed with ED information.

\begin{figure}[t]
\begin{center}
\psfig{file=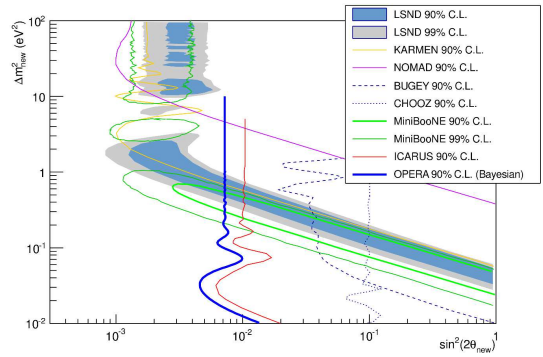,width=2.8in}
\end{center}
\caption{The constraints on the LSND/MiniBooNE anomaly set by the OPERA experiment together with the constraints from the other experiments (see \cite{aga13d} for references).}
\label{fig:fig5}
\end{figure}

It was possible to set a bound on $sin^{2}(2\theta_{13})< 0.44$ at the 90 \% C. L \cite{aga13d}. The constraints set on the LSND/MiniBooNE anomaly \cite{aqu01,aqu12} parameters are presented in fig.~\ref{fig:fig5}. All results obtained so far are compatible with the hypothesis that all observed $\nu_{e}$ events are due to prompt component and other sources of background (the latter was estimated as 0.4$\pm$0.2 events).

\section{Conclusions} \label{sec:conc}
We have reviewed the present status of the OPERA experiment aimed on event-by-event $\nu_{\mu} \rightarrow \nu_{\tau}$ oscillation search in appearance mode. Three $\nu_{\tau}$ candidates were observed so far. The main sources of background were carefully studied. The complete simulation/analysis framework OpRelease was recently established, allowing to perform more realistic MC studies. Constraints on $\theta_{13}$ and the LSND/MiniBooNE anomaly were set profiting of the excellent particle identification capabilities of the OPERA detector.

\section*{Acknowledgements} \label{sec:ackn}
Work of T. Dzhatdoev was partially supported by the Russian Foundation for Basic Research, grant 12-02-12142-ofi\_m.

\end{document}